# MESSAGE ROUTING IN WIRELESS AND MOBILE NETWORKS USING TDMA TECHNOLOGY

Timotheos Aslanidis[1] and Leonidas Tsepenekas[2]

[1]National Technical University of Athens, Athens, Greece
taslan.gr@gmail.com
[2]National Technical University of Athens, Athens, Greece
ltsepenekas@corelab.ntua.gr

## ABSTRACT

*In an era where communication has a most important role in modern societies, designing efficient algorithms for data transmission is of the outmost importance. TDMA is a technology used in many communication systems such as satellite, cell phone as well as other wireless or mobile networks. Most 2G cellular systems as well as some 3G are TDMA based. In order to transmit data in such systems we need to cluster them in packages. To achieve a faster transmission we are allowed to preempt the transmission of any packet in order to resume at a later time. Preemption can be used to reduce idleness of some stations. Such preemptions though come with a reconfiguration cost in order to setup for the next transmission. In this paper we propose two algorithms which yield improved transmission scheduling. These two algorithms we call MGA and IMGA (Improved MGA). We have proven an approximation ratio for MGA and ran experiments to establish that it works even better in practice. In order to conclude that MGA will be a very helpful tool in constructing an improved schedule for packet routing using preemtion with a setup cost, we compare its results to two other efficient algorithms designed by researchers in the past: A-PBS(d+1) and GWA. To establish the efficiency of IMGA we ran experiments in comparison to MGA as well as A-PBS(d+1) and GWA. IMGA has proven to produce the most efficient schedule on all counts.*

## 1. INTRODUCTION

In the course of the last fifty years technological and scientific evolution has lead to an era of vast information and the need for fast and efficient communication. In the framework of enhancing communication network performance and dissemination of information researchers have introduced the Time Division Multiple Access (TDMA) technology. TDMA technology has been for decades a cornerstone of the global network infrastructure, as it plays an important role in many different communication systems. We can abstractly describe the TDMA technology with the following mathematical model. There are N transmitters and M receivers connected in a NxM switching board. At any time the switching board has a state, which is defined by pairs of connected transmitters and receivers. At a given state each transmitter is connected to only one receiver and vice versa. This state of connections allows the transmission of data between the agents of each pair. Prior to the start of the whole process the amount of information (with known transmission time) that needs to be transmitted between each transmitter-receiver pair is known. Also, every state of the switching board defines a package of data (a phase of the transmission) that will be transmitted. The time overhead of the above described package is characterized by the time needed for the longest transmitter-receiver pair included in the package. Another critical feature of these systems that we need to take into account is that every time the switching board is reconfigured there is an additional time cost. All the above can be better understood through the following example:

Suppose we have a 3x3 system with the traffic matrix shown in figure 1 (senders in rows – receivers in columns). Every cell (i, j) of the traffic matrix contains the time needed for the transmission of all the data between sender i and receiver j. Also, suppose that the cost of reconfiguring the board is 2 time units. Then figure 2 shows the whole scheduling procedure.

|   |    |    |
|---|----|----|
| 7 | 12 | 9  |
| 8 | 11 | 14 |
| 0 | 9  | 13 |

Figure 1. Example of a traffic matrix

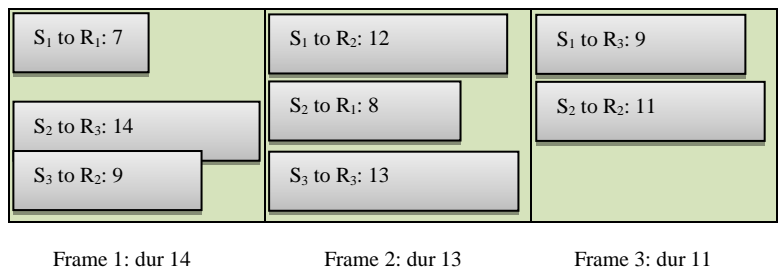

Frame 1: dur 14      Frame 2: dur 13      Frame 3: dur 11

Figure 2. A feasible schedule for corresponding to the traffic matrix in figure 1

In the above example the total transmission time is $14 + 13 + 11 + 3 \times 2 = 44$ time units. The duration of the first package/frame is dominated by the transmission time of the message between sender 2 and receiver 3 and will therefore need 14 time units. The same applies to the following 2 frames as well. We also pay $3 \times 2$ because we need to reconfigure the board three times.

TDMA based systems aim in transmitting data between multiple sender and receiver stations in packages simultaneously. While trying to reduce the total time frame, preemption of a transmission is allowed in order to send the remaining parts of the messages at a later time with a newly scheduled package. If we take preemption into account, the setup cost will influence the overall makespan significantly, and this is exactly why the optimal solution is so hard to calculate. This problem is referred to in the literature as the MINSWT problem in case the number of frequencies does not suffice to serve all stations at once. In case the number of frequencies is at least as large as the number of senders as well as the number of receivers the problem is referred to as PBS. In this paper we handle the later.

Here are some examples of networks that are based on the aforementioned TDMA technology.

- Most 2G cellular systems are TDMA based. The GSM (Global System for Mobile Communications) currently accounts for approximately 80% of the subscribers worldwide. Many other 2G systems use TDMA technology among which are Personal Digital Cellular (PDC), the Digital Enhanced Cordless Telecommunications (DECT) standard for portable phones and PHS. Surprisingly enough 2G systems are not at all obsolete. They are still used often used independently or in co-existence with the newest 3G and 4G systems.

- TDMA technology is also used in some 3G cellular systems such as the Universal Mobile Telecommunications System (UMTS).
- TDMA technology is also still used in satellite systems, in combat-net radio systems and in the Passive Optical Networks (PON).

## 2. GRAPH REPRESENTATION AND NOTATIONS

For the purposes of our research we will represent an input instance by a bipartite graph $G(V,U,E,w)$. V will denote the transmitters, U will stand for receivers, whereas the set of edges will comprise the information about data traffic through the TDMA system. The weight $w(v,u)$, assigned to each edge $e=(v,u)$, $v \in V$, $u \in U$ is the time required for the full transmission of each message.

Furthermore the following notation will be used: $\Delta = \Delta(G) = \max\{\max_{v \in V}(\deg(v)), \max_{u \in U}(\deg(u))\}$, that is, $\Delta$ will denote degree of the bipartite graph which in practice equals to the maximum number of messages to be transferred from or to any of the stations.

$W = W(G) = \max\{\max_{v \in V}\{\sum_{u \in U} w(v,u)\}, \max_{u \in U}\{\sum_{v \in V} w(v,u)\}\}$, that is W will denote the maximum total weight of all the edges adjacent to any of the nodes. This in turn equals to the maximum total workload of any station.

$d \in Z^*_+$ will denote the setup delay, namely the time required so that the next transmission may begin.

The objective function to be minimized is $F(G,d) = \sum_{i=1}^{N} t(M_i) + d \cdot N$, where N is the number of distinct transmissions (phases) in order to transfer the entire data workload and $t(M_i)$ is the time required for the completion of a specific transmission $M_i$.

Since transmission cannot be concluded before the maximum workload of any station is scheduled and the number of transmissions will be at least as many as the messages to be sent or received by any station, a lower bound to the optimal solution is $LB = W + d \cdot \Delta$. Yet, this lower bound is not always achievable as shown in [6].

## 3. PREVIOUS RESEARCH

As shown in [4], PBS is $4/3-\varepsilon$ inapproximable for any $\varepsilon > 0$, unless P=NP. Even though the problem is NP-Hard there do exist special cases of input for which the optimal solution can be found in polynomial time ([1], [4], [5], [6]). The best approximation ratio proven so far is $2 - \frac{1}{d+1}$ by the authors of [1]. Experiments have been ran by many researchers to test the output of various algorithms proposed in [2], [4], [5], [6], [10] and [12].

The performance of our newly presented algorithm will be compared to that of two algorithms found in bibliography:

- The algorithm presented in [8] which we will refer to as GWA (Gopal-Wong Algorithm). GWA calculates exactly Δ matchings, corresponding to Δ transmission packages. GWA will always achieve the minimum number of switchings and in order to produce a competitive transmission time for each package, the matchings are constructed so that edges of similar weight are grouped together. GWA has been tested in experiments in [6] and appears to perform well when the value of d increases significantly compared to duration of the messages. Unfortunately it has an unbounded approximation ratio as shown also in [6].

- A-PBS(d+1) as described in [1], preempts each edge to a multiple of d+1 and repeatedly computes matchings that correspond to transmission packets. Until now A-PBS(d+1) is the only algorithm that has a proven approximation ratio strictly less than 2. Yet, in most cases it produces schedules with makespan undesirably larger than the optimal.

Our newly developed algorithm, which we call MGA aims in mitigating these disadvantages of WGA and A-PBS(d+1). MGA tackles GWA's disadvantage, namely the fact that there are instances for which GWA produces a solution of unbounded approximation ratio and in addition it produces schedules that are on average a lot close to the optimal than those produced by A-PBS(d+1). Table 1 summarizes all of the above.

Table 1. Summary of the 3 algorithms comparison: GWA, A-PBS(d+1), MGA

| Algorithm | Approximation ratio | Experimental results' conclusions |
|---|---|---|
| GWA | Unbounded | Works well only for large values of d and works undesirably bad for specific instances regardless the value of d. |
| A-PBS(d+1) | $2 - \dfrac{1}{d+1}$ | Often produces results with more than 50% deviation from the optimal. |
| MGA | Δ+1 | Produces efficient schedules on average as well in the worst case scenario regardless the input. |

## 4. MGA: AN IMPROVED ROUTING ALGORITHM FOR DATA TRANSMISSION IN TDMA SYSTEMS

For the purposes of this paper we have designed an algorithm aiming in mitigating the disadvantage of GWA, namely an algorithm with a bounded approximation ratio. We will refer to this algorithm as MGA (MultiGraph Algorithm), as the main concept in order to achieve a bounded approximation ratio is to split each edge of undesirably large weight into smaller edges to be handled and scheduled independently.

*The MultiGraph Algorithm (MGA)*

*Step1:* Split each edge of weight more than $\left\lfloor \dfrac{W}{\Delta} \right\rfloor$ in parts each having weight no more than $\left\lfloor \dfrac{W}{\Delta} \right\rfloor$. The splitting will be done in the following way: Split each edge $e \in E$ with weight w(e) into at most $\left\lfloor \dfrac{w(e)\Delta}{W} \right\rfloor + 1$ edges the weight of each of which will be $\left\lfloor \dfrac{W}{\Delta} \right\rfloor$ except perhaps for the last one which will weigh $w(e)-\lfloor\frac{w(e)\Delta}{W}\rfloor\cdot\lfloor\frac{W}{\Delta}\rfloor = w(e)$ MOD $\lfloor\frac{W}{\Delta}\rfloor$. Thus G will become a multigraph.

*Step 2:* Add nodes and edges to the multigraph in order to make it a regular multigraph. Each newly added edge e, will have w(e)=0.

*Step 3:* Compute a perfect matching for the regular multigraph and schedule the corresponding parts of the edges of this matching for transmission.

*Step 4:* Remove the edges corresponding to the previous transmission from the multigraph.

*Step 5:* repeat steps 3 and 4 until E=∅.

*Theorem1:* MGA's approximation ratio is bounded by Δ+1.

*Proof:* In the multigraph constructed by steps 1 and 2 the maximum edge weight is $\lfloor\frac{W}{\Delta}\rfloor$. Therefore the cost of each transmission will not exceed $\lfloor\frac{W}{\Delta}\rfloor$. The multigraph's degree is at most $\Delta' \leq (\lfloor\frac{w_{max}\cdot\Delta}{W}\rfloor+1)\cdot\Delta$ since there can be at most Δ edges to be split and each will be split in at most $\lfloor\frac{w_{max}\cdot\Delta}{W}\rfloor+1$ parts, where $w_{max}$ is the maximum weight of any edge in the graph. Step 2 ensures that each node has degree Δ′ and that removing the edges of a perfect matching from G will reduce the graph's degree by exactly one after each iteration. Thus the number of iterations will be Δ′. Therefore the cost C of the entire process to transmit all data will be bounded by:

$$C \leq (\lfloor\frac{w_{max}\cdot\Delta}{W}\rfloor+1)\cdot\Delta\cdot\lfloor\frac{W}{\Delta}\rfloor+d\cdot\Delta' \leq (\lfloor\frac{w_{max}\cdot\Delta}{W}\rfloor+1)\cdot\Delta\cdot\lfloor\frac{W}{\Delta}\rfloor+d\cdot(\lfloor\frac{w_{max}\cdot\Delta}{W}\rfloor+1)\cdot\Delta$$

Taking into account that $w_{max}\leq W$, $\lfloor a \rfloor \leq a$, for all $a\in Q$ we conclude that

$$C \leq (\frac{W\cdot\Delta}{W}+1)\cdot\Delta\cdot\frac{W}{\Delta}+d\cdot(\frac{W\cdot\Delta}{W}+1)\cdot\Delta = (\Delta+1)\cdot W+d\cdot(\Delta+1)\cdot\Delta = (\Delta+1)\cdot(W+d\cdot\Delta)$$

Which implies that C≤(Δ+1)·LB, thus bounding MGA's approximation ratio by Δ+1.

## 5. RUNNING TEST CASES TO EVALUATE THE PERFORMANCES OF THE THREE ALGORITHMS

One hundred test cases have been ran for a 50 source-50 destination system for values of setup cost varying from 1 to 100 and message durations varying from 1 to 200. We have to point out that since PBS is an NP-Hard problem, calculating an optimal schedule is inefficient therefore to estimate the approximation ratio we have used the lower bound to the optimal solution which is W+Δ·d.

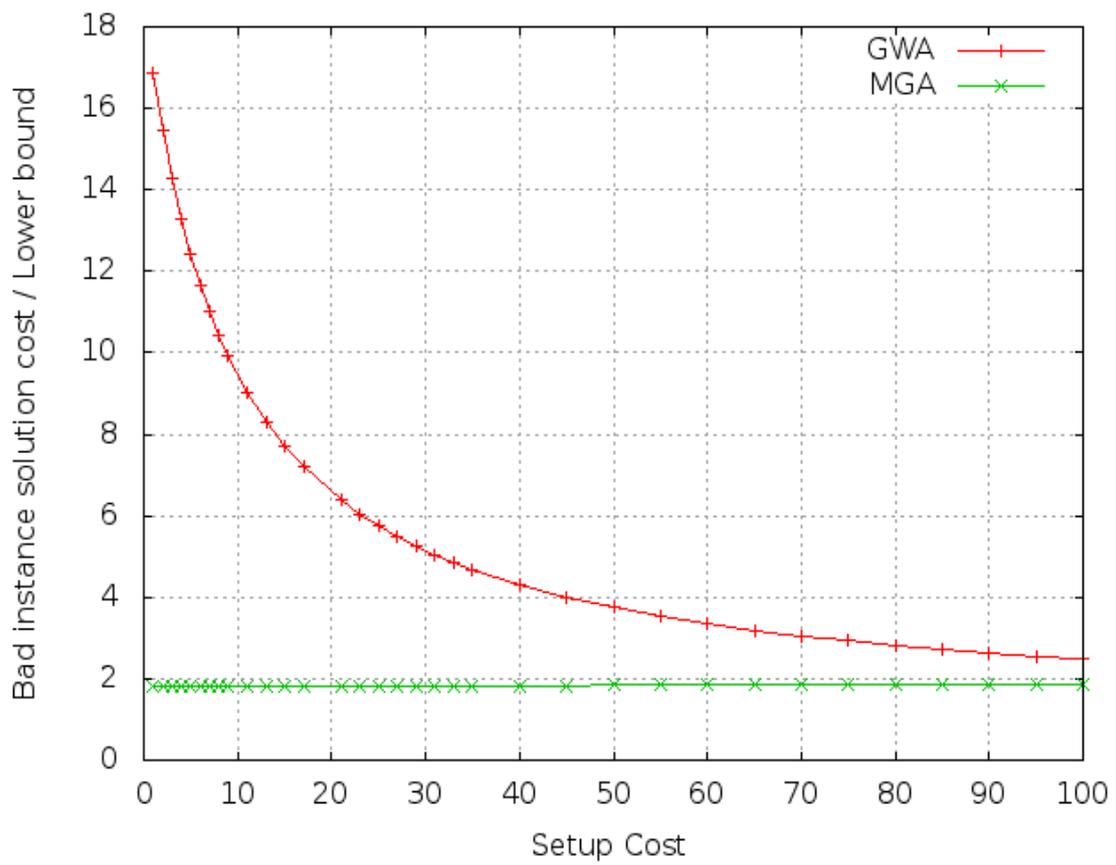

Figure 3. Solution cost/lower bound comparison of GWA and MGA for a single "bad" instance.

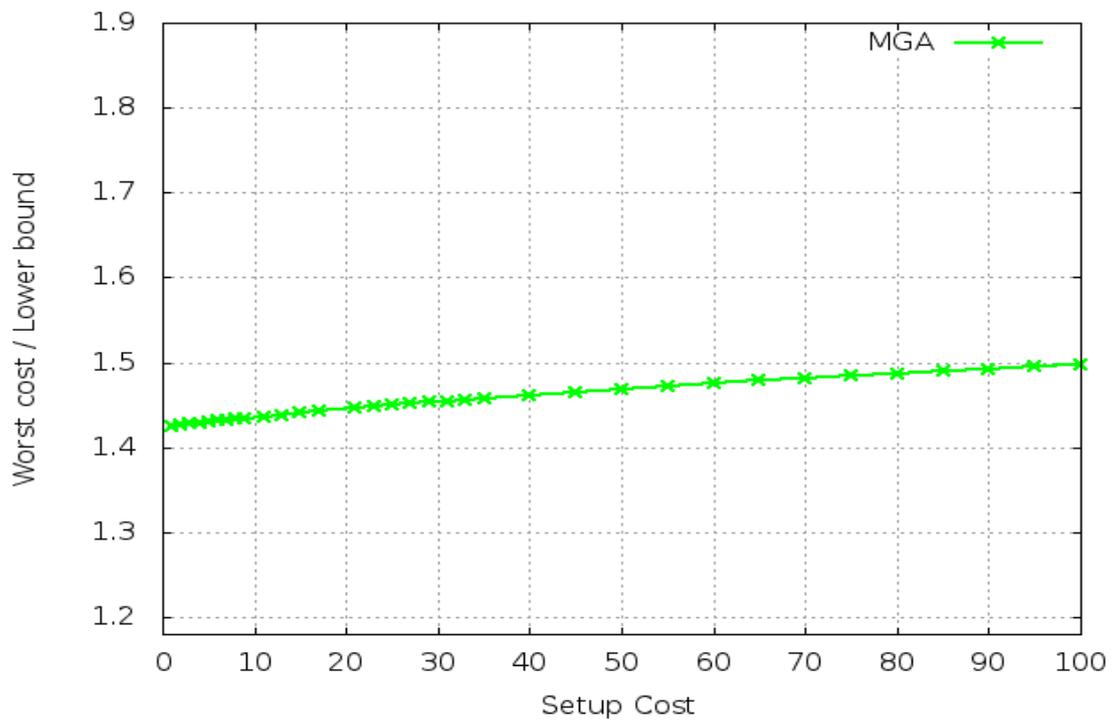

Figure 4. worst performance of MGA

Figure 3 establishes that MGA works better than GWA not only in the theoretical sense that theorem 1 implies but also in practice as well. We ran both algorithms using as input the "bad" instance presented in [6]. MGA will still yield an approximation ratio lower than 2 and will regardless the value of d perform better than GWA. We thereof have established that our newly presented algorithm will perform well, even for the worst transmission scenario.

Figure 4 presents the worst performance of MGA. It suggests that even though our proven approximation is Δ-dependent, in practice MGA will not exceed an approximation ratio of 2 or even less. In fact MGA's (worst case/lower bound) will in no case exceed 1.55. Furthermore MGA's worst performance for any instance does not fluctuate much from its average performance, making it a stable and reliable tool for constructing an efficient schedule for the problem at hand.

Figure 5 compares MGA with A-PBS(d+1). A-PBS(d+1) will perform better only for very small values of d and even though it has a better approximation ratio, MGA produces a lot better results as d's value increases.

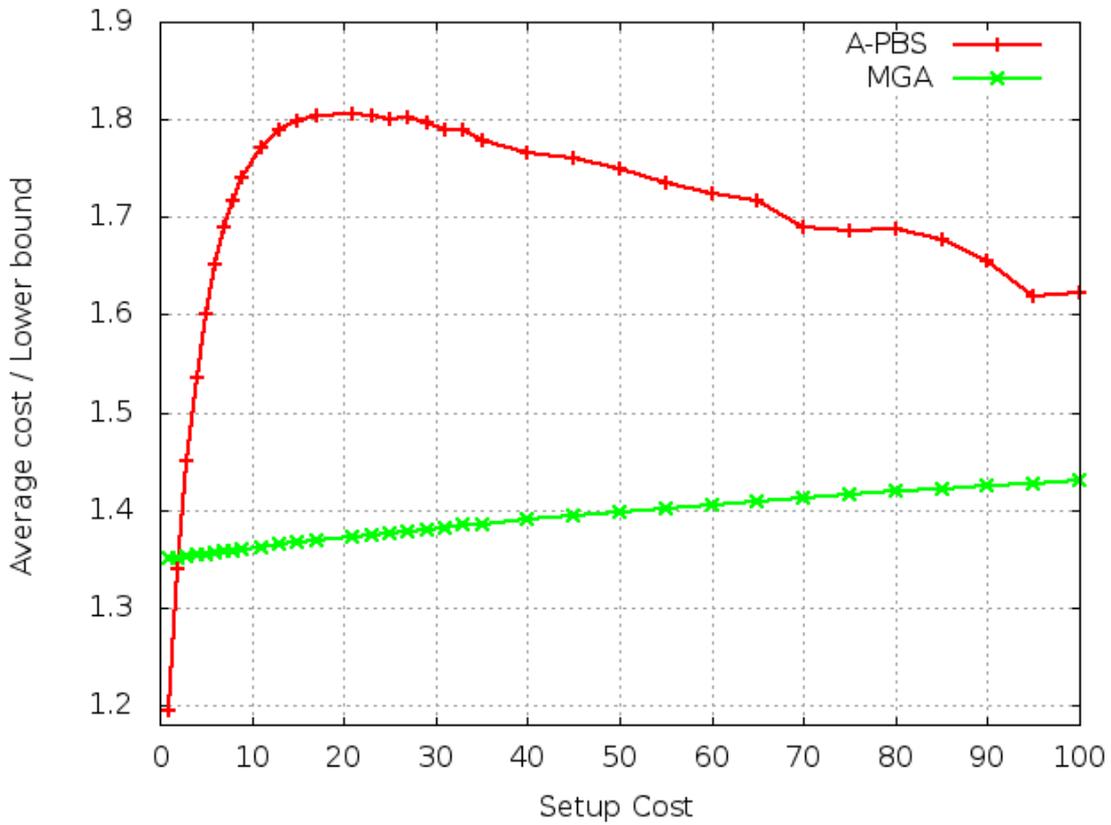

Figure 5: Average cost/lower bound comparison of A-PBS(d+1) and MGA.

## 6. COMPLEXITY COMPARISON OF THE THREE ALGORITHMS

To calculate the computational complexity of the 3 algorithms we assume that the initial graph is G(U,V,E) with max{|U|, |V|}=n and |E|=m.

GWA's complexity: In this algorithm a subroutine is used to turn the original graph into a regular graph. This subroutine adds nodes and edges as needed, with a running time up to $O(n^2)$. A sorting of the edges according to their weight follows. The cost for this sorting is $O(n\log n)$. Finally the authors propose a special way of calculating a perfect matching in an attempt to reduce the messages transmission time (subject to the number of preemptions being minimum) which will need $O(n^3)$ steps to conclude. The number of these matchings will be $O(n)$, yielding a total of $O(n^4)$ complexity to decide how to schedule the packets. The overall complexity of GWA is therefore $O(n^2)+O(n\log n)+O(n^4)=O(n^4)$.

The complexity of A-PBS(d+1) as calculated in [12] is $O(\sqrt{n} \cdot m^2)$.

MGA's complexity is determined by the perfect matching algorithm used to be multiplied with the number of times that the perfect matching algorithm will run. Using the algorithm in [13] each matching costs $O(\sqrt{n} \cdot m)$ and as shown in the proof of theorem1 the number of such matchings will not exceed $\Delta(\Delta+1)=O(n^2)$. Consequently the overall complexity of MGA is $O(n^2\sqrt{n} \cdot m)$.

Depending on the actual value of |E|=m which will vary from $O(n)$ to $O(n^2)$, the faster of the three algorithms would be either GWA or A-PBS(d+1). Yet, in the test cases ran, MGA will more often than not, calculate no more than $O(n)$ perfect matchings (instead of $O(n^2)$ that are expected in the worst case) thus improving its actual running time by $O(n)$ and yielding the requested schedule by up to $O(n^{0.5})$ faster than any of the other two algorithms considered.

## 7. IMPROVING MGA'S PERFORMANCE

Let G(U,V,E) denote the initial graph. We consider the following heuristic variation of MGA:

*Improved MultiGraph Algorithm (IMGA)*

Step1: Add nodes and edges to G(V,U,E) to make it regular. Each newly added edge e will have w(e)=0. Denote the regular graph by $G_1(V_1,U_1,E_1)$.
Step2: Construct $G_2(V_2,U_2,E_2)$ from G(V,U,E) in the same way as described in step1 of MGA. For any u∈U and v∈V with (u,v)=e let $w_{max}(u,v)=w_{max}(e)$, denote the maximum weight of each of the split edges of the induced graph.
Step3: Use subroutine 1 to calculate a perfect matching M in $G_1$.
Step4: For each edge e∈M, reduce its weight in G by $w_{max}(e)$.
Step5: Repeat steps 1 to 3 until there are no more edges left in E.

Subroutine1:
Step1: Sort the edges by decreasing order of weight. Let L={$e_1, e_2, …, e_{|E|}$} be the induced list.

Step2: P←{$e_1$}, M←{$e_1$}, i←1.

Step3: Set P←P∪{$e_i$}. Search for an augmenting path for M in P. If such a path exists then augment M along P.

Step4: i←i+1

Step5: Repeat steps 3 and 4 until M is perfect.

Subroutine1 is the same as proposed in [8]. Only difference is that for the purposes of this manuscript the edges are in decreasing order of weight. Proof that it calculates a maximum matching in a regular graph can be found in [8] as well.

Here is an example of how IMGA might yield better results than MGA:
Consider the following traffic matrix:

$$A = \begin{pmatrix} 7 & 3 & 4 \\ 3 & 4 & 3 \\ 0 & 5 & 4 \\ 2 & 8 & 5 \end{pmatrix}$$

Each row corresponds to a sender and each column to a receiver. The entries stand for the transmission times. W=20 and Δ=4. Therefore $\left\lfloor \dfrac{W}{\Delta} \right\rfloor$ =5 and splitting the edges will result in the following traffic matrix:

$$A_1 = \begin{pmatrix} 5+2 & 3 & 4 \\ 3 & 4 & 3 \\ 0 & 5 & 4 \\ 2 & 5+3 & 5 \end{pmatrix}$$

A possible schedule produced by MGA will comprise 5 transmissions which might be the following

$$T_1 = \begin{pmatrix} 5 & 0 & 0 \\ 0 & 0 & 0 \\ 0 & 0 & 4 \\ 0 & 3 & 0 \end{pmatrix}, T_2 = \begin{pmatrix} 2 & 0 & 0 \\ 0 & 0 & 3 \\ 0 & 5 & 0 \\ 0 & 0 & 0 \end{pmatrix}, T_3 = \begin{pmatrix} 0 & 0 & 4 \\ 3 & 0 & 0 \\ 0 & 0 & 0 \\ 0 & 5 & 0 \end{pmatrix}, T_4 = \begin{pmatrix} 0 & 3 & 0 \\ 0 & 0 & 0 \\ 0 & 0 & 0 \\ 0 & 0 & 5 \end{pmatrix}, T_5 = \begin{pmatrix} 0 & 0 & 0 \\ 0 & 4 & 0 \\ 0 & 0 & 0 \\ 2 & 0 & 0 \end{pmatrix}.$$

The time required to complete data transmission using this schedule produced by MGA will be 24+5d, since t($T_1$)=5, t($T_2$)=5, t($T_3$)=5, t($T_4$)=5 and t($T_5$)=4.

IMGA will yield a much better schedule since step 3 of the algorithm tries to group together the longest messages that are to be transmitted just as they were in the initial data. Here is one such schedule:

$$T_1 = \begin{pmatrix} 5 & 0 & 0 \\ 0 & 0 & 0 \\ 0 & 0 & 4 \\ 0 & 5 & 0 \end{pmatrix}, T_2 = \begin{pmatrix} 0 & 0 & 0 \\ 3 & 0 & 0 \\ 0 & 5 & 0 \\ 0 & 0 & 5 \end{pmatrix}, T_3 = \begin{pmatrix} 0 & 0 & 4 \\ 0 & 4 & 0 \\ 0 & 0 & 0 \\ 2 & 0 & 0 \end{pmatrix}, T_4 = \begin{pmatrix} 0 & 3 & 0 \\ 0 & 0 & 3 \\ 0 & 0 & 0 \\ 0 & 3 & 0 \end{pmatrix}, T_5 = \begin{pmatrix} 2 & 0 & 0 \\ 0 & 0 & 0 \\ 0 & 0 & 0 \\ 0 & 0 & 0 \end{pmatrix}.$$

The time required to complete data transmission using this schedule produced by IMGA will be 19+5d, since $t(T_1)=5$, $t(T_2)=5$, $t(T_3)=4$, $t(T_4)=3$ and $t(T_5)=2$.

To further illustrate how IMGA produces improved schedules compared to MGA we ran test cases, the results of which are depicted in figures 6 to 8. We ran one hundred test cases for a 50 source-50 destination system for values of setup cost varying from 1 to 100 and message durations varying from 1 to 200.

Figure 6 shows how IMGA performs in average compared to MGA. IMGA may provide results with an approximation ratio by up to 10% improved compared to that of MGA as the value of d increases.

Figure 7 compares the average performance of IMGA with A-PBS(d+1) and GWA. IMGA has a somewhat stable average approximation ratio, regardless the value of d. Compared to A-PBS(d+1) and GWA, it produces a schedule closer to the lower bound for all values of d tested.

Figure 8 shows the worst approximation ratio of IMGA and compares it to that of MGA. The worst approximation ratio that IMGA achieves is by up to 9% better than that of MGA, as the reconfiguration cost increases.

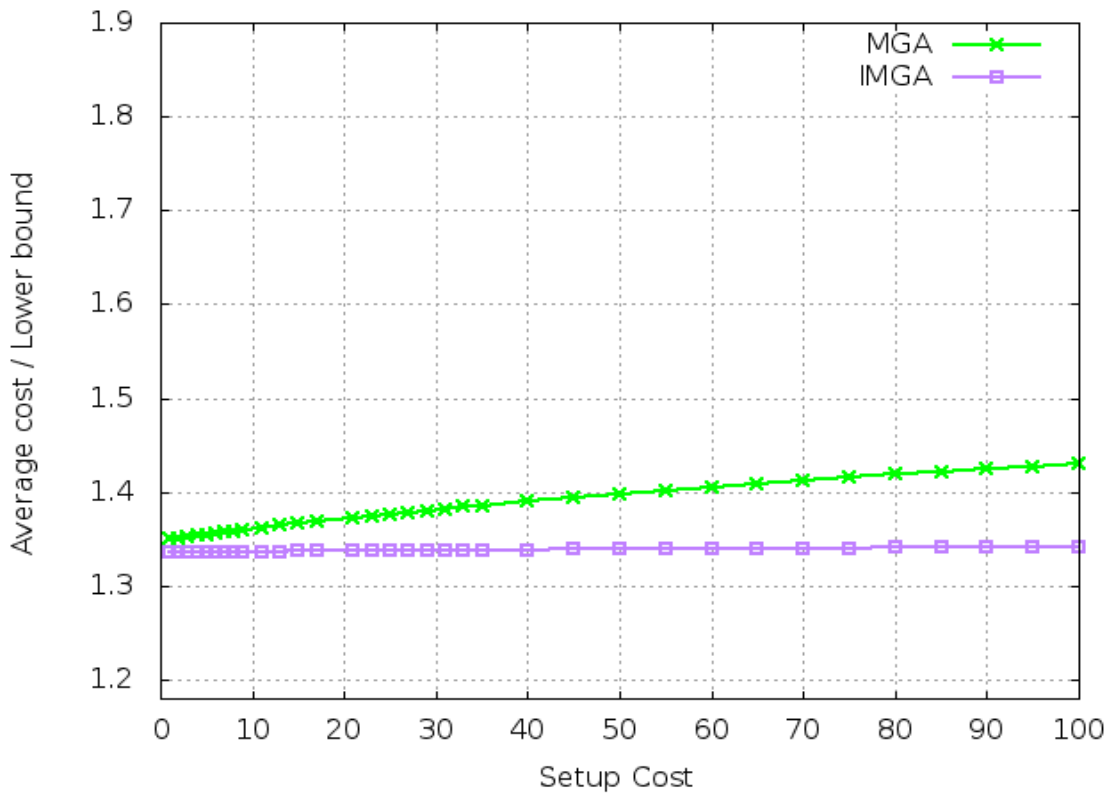

Figure 6: Average cost/lower bound comparison of IMGA and MGA.

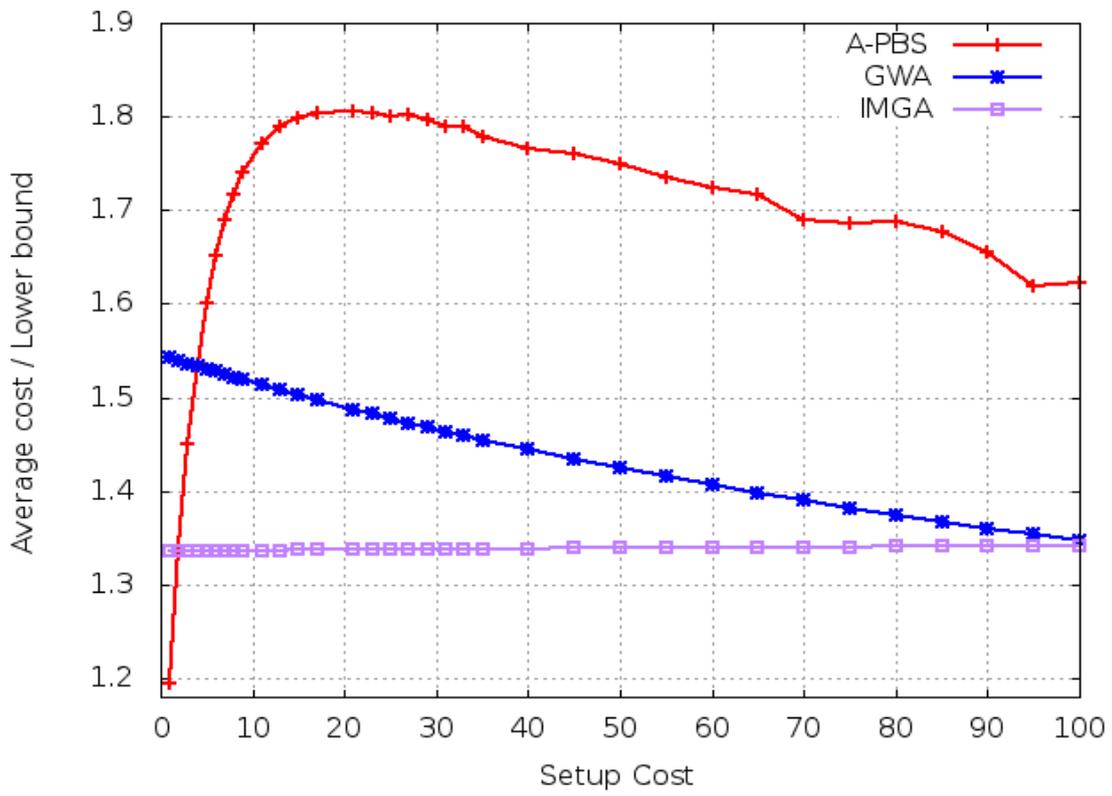

Figure 7: Average cost/lower bound comparison of IMGA, A-PBS(d+1) and GWA.

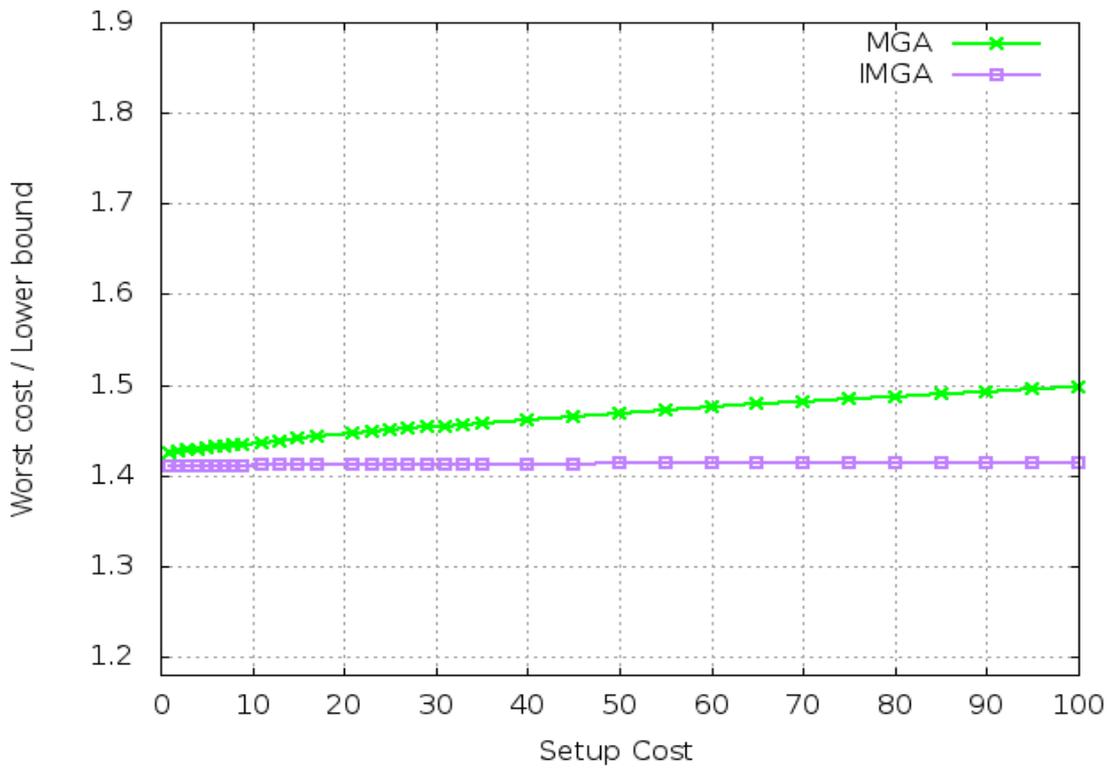

Figure 8: worst cost/lower bound comparison of IMGA and MGA.

# 8. CONCLUSIONS AND FUTURE WORK

In this paper we have presented MGA, a Δ+1- approximation algorithm for the problem of transmitting data packages through a TDMA based wireless or mobile communication system. Furthermore, we ran experiments to establish how efficient MGA is in practice. Experiments suggest that it might be possible to prove a better approximation ratio than Δ+1. That approximation ratio may even be less than two. We compared MGA with two algorithm found in bibliography. One which achieved the minimum number of preemptions and another which has the best approximation ratio proven so far, to establish that MGA works even better in practice. Yet, in order to prove MGA's approximation ratio we designed MGA so that it forcefully preempts transmission numerous times resulting in a schedule burdened by many delays. Future work might also suggest of a way to reduce the number of those preemptions leading to even better experimental results or even a proof for a lower approximation ratio. In order to further improve on MGA's results we proposed a heuristic variation of MGA, which we call IMGA. IMGA yields even better schedules and even though we have no proof of it yet, is seems to also have a bounded approximation ratio.

**Authors**

Timotheos Aslanidis was born in Athens, Greece in 1974. He received his Mathematics degree from the University of Athens in 1997 and a master's degree in computer science in 2001. He is currently doing research at the National and Technical University of Athens in the School of Electrical and Computer Engineering. His research interests comprise but are not limited to computer theory, number theory, network algorithms and data mining algorithms.

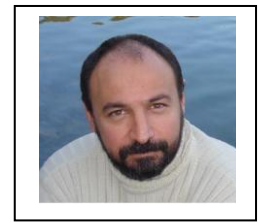

Leonidas Tsepenekas was born in Athens, Greece in 1992. Currently, he is finishing his studies as an undergraduate student at the National Technical University of Athens. His main research interests focus on approximation, online and randomized algorithms.

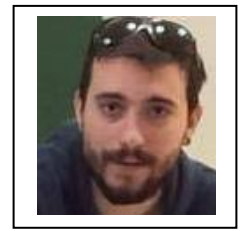